\documentclass[aps,nofootinbib]{revtex4}

\usepackage{amsmath}
\usepackage{amsfonts,amssymb,amsthm, bbm,braket}
\usepackage{graphicx}   
\usepackage{subfigure}  
\usepackage{amsbsy} 
\usepackage[bold]{hhtensor} 
\usepackage{algpseudocode}
   
\usepackage{multirow}

\newcommand\Tr{\mathrm{Tr}}

\usepackage{hyperref} 

\hypersetup{
  colorlinks   = true, 
  urlcolor     = blue, 
  linkcolor    = blue, 
  citecolor   =  red 
}

\begin{document}
\title{Quantum Model Averaging}

\author{Christopher Ferrie}
\affiliation{
Center for Quantum Information and Control,
University of New Mexico,
Albuquerque, New Mexico, 87131-0001, USA}

\date{\today}


\begin{abstract}
Standard tomographic analyses ignore model uncertainty.  It is assumed that a given model generated the data and the task is to estimate the quantum state, or a subset of parameters  within that model.  Here we apply a model averaging technique to mitigate the risk of overconfident estimates of model parameters in two examples: (1) selecting the rank of the state in tomography and (2) selecting the model for the fidelity decay curve in randomized benchmarking.
\end{abstract}


\maketitle
\tableofcontents

\section{Introduction}

Parameter estimation is an integral part of physics.  Accurate estimates of physical parameters in quantum mechanical models allows for precision quantum control \cite{khaneja2005optimal, brif_2010_control} which enables practical goals such as quantum computation \cite{nielsen2010quantum} and quantum metrology \cite{caves1981quantum} which in turn can provide probes of fundamental physics such as gravity wave detection \cite{aasi2013enhanced}.

Quantum parameter estimation shares many similarities with its classical counterpart.  But there are many subtle and peculiar differences.  Even in single parameter estimation, {quantum metrology} \cite{Helstrom1976Quantum,caves1981quantum} shows that we can obtain advantages from quantum resources such as squeezed states of light \cite{Lang2013Optimal} and entanglement \cite{giovannetti_2006_quantum}.  With such rich structure, many new subtleties \cite{barndorff2000fisher}, considerations \cite{fujiwara_2008_fibre, knysh_2011_scaling, escher_2011_general} and generalities \cite{Boixo2007Generalized,Tsang2011Fundamental,ferrie_2014_Fisher} can arise, including entirely new approaches to estimation \cite{Higgins2007Entanglement,Hentschel2011Efficient,mahler2013adaptive,ferrie_likelihood-free_2014} and verification \cite{flammia2011direct, da2011practical, magesan_efficient_2012}.  

At the other end of the parameter spectrum the quantum state of a physical system is our most complete description of it. Thus estimation of quantum states \cite{paris2004quantum} might be seen as the most ambiguous form of estimation.  There are many approaches to the general problem \cite{hradil1997quantum,blume2010optimal, blume2010hedged,gross2010quantum}, some which specialize for computational efficiency \cite{Gottesman2008Identifying, Cramer2010Efficient,Toth2010Permutationally,LandonCardinal2012Practical} and those which go beyond to estimation of regions \cite{Rehacek2008Tomography, Christandl2011Reliable,Arrazola2013Reliable, BlumeKohout2012Robust,Shang2013Optimal,ferrie_high_2013}.

All of the above mentioned results assume a model.  That is, it was taken as given that a particular parametric distribution generated the data.  But what if this assumption is not correct?  This question has garnered recent interest and classical approaches to \emph{model selection} have been used in a variety of experimental \cite{usami_2003_accuracy,guta_2012_rank} and theoretical \cite{lougovski_2009_characterizing, lin_2011_information, schwarz_2013_error, vanenk_2013_when} works.  Here we supplement these results with a new approach to parameter estimation: \emph{model averaging}.  The technique shares many similarities with model selection---in fact, model selection is a crucial component of model averaging, but goes beyond it.  Although model selection adds an additional layer of security to overconfident estimation, selecting models can be itself a red herring, for the most probable model might only be slightly more probable than others.  

The approach considered here combines Bayesian parameter estimation with Bayesian model selection, such that the final estimate of the parameters is the best value of the parameters within each model, averaged over the probability assigned to each model.  We will show that such an approach can reduce the error incurred by first selecting a model---which has some probability of being incorrect---then selecting parameters within that model.  In fact, the numerical experiments presented here show that model average estimation always does better than the estimates from incorrect models and in some scenarios can perform better than even estimates from the correct model.  This is due to the additional hedging afforded by considering multiple models, all of which carry \emph{some} information.

The paper is organized as follows.  In Sec.~\ref{sec:review} we outline the problem and review the model selection techniques used so far.  In Sec.~\ref{sec:Bayes}, we present the full Bayesian approach to model selection and define the \emph{model average estimate} of parameters.  Sec.~\ref{sec:examples} presents two distinct examples where the model average estimate provides an advantage for parameter estimation.  We conclude with a discussion in Sec.~\ref{sec:conclusion}.

\section{General problem and common methods \label{sec:review}}

In Sec. \ref{sec:A} we overview the problem.  In Secs. \ref{sec:AIC} and \ref{sec:BIC} we review the Akaike information criterion (AIC) and the Bayesian information criterion (BIC) which are by far the most commonly used approaches to model selection (and the only used thus far for quantum estimation).  These sections are included for reference and completeness.

\subsection{Problem setup \label{sec:A}}
Let us begin with the base problem of \emph{parameter estimation}.  A physical model prescribes the probabilities for the outcomes of experiments: $\Pr(D|\vec{x};C,M)$.  Here $D$ is some hypothetical or observed data, $\vec{x}$ is a set of real numbers in $\mathbb R^d$, where $d$ is the \emph{dimension} of the model, $C$ is the experimental context\footnote{Experimental context, $C$, here means any additional information necessary to provide a well defined function $\Pr(D|\vec{x};C,M)$.} and $M$ is our model.  For example, we could have the model $M$ be that of qubit which is parameterized by a Bloch vector $\vec{x} = (x,y,z)$.  The experimental context could be a measurement basis, say that of $\sigma_x$.  Then, the quantum mechanical model prescribes $\Pr(\pm|\vec{x};C,M) = (1\pm x)/2$.  In quantum mechanics this is called the \emph{Born rule} and in statistics, the \emph{likelihood function}.  

Going from parameters to the probability of data is a deductive process---the model gives us numerical values of $\Pr(D|\vec{x};C,M)$.  The experiment, on the other hand, gives us a particular data set $D$---but what we really want is $\vec{x}$.  This is an example of an \emph{inverse problem}.  What the Bayesian solution provides is $\Pr(\vec{x}|D;C,M)$, the \emph{distribution} of $\vec{x}$ given the data.  Although we lack certainty about $\vec{x}$, we can accurately (read: quantitatively, mathematically) describe our state of knowledge of $\vec{x}$ given we have seen the data $D$. 

The formal path forward is through Bayes' rule:
\begin{equation}
\Pr(\vec{x}|D;C,M) = \frac{\Pr(D|\vec{x};C,M)\Pr(\vec{x}|C,M)}{\Pr(D|C,M)}.
\end{equation}
From a chronological point of view relative to $D$, we begin with the \emph{prior} $\Pr(\vec{x}|C,M)$, which encodes the information we have about $\vec{x}$ prior to learning about $D$.  We weight the prior by the likelihood function and normalize by the \emph{marginal likelihood}\footnote{The notation $\mathbb E_x[f(x)]$ means take the expectation of $f(x)$ with respect to the distribution of $x$ (the distribution itself is implicit).}:
\begin{equation}\label{marginal likelihood}
\Pr(D|C,M) = \mathbb E_{\vec{x}|C,M}[\Pr(D|\vec{x};C,M)].
\end{equation}
The distribution produced as result of Bayes rule, $\Pr(\vec{x}|D;C,M)$, is called the \emph{posterior} and represents our knowledge of $\vec{x}$ after the data has been observed.

At this point we could call the problem solved.  This, however, assumes the model is correct.  If the model is suspect, then we have the meta--problem of determining the best model.

\subsection{Akaike Information Criterion (AIC)\label{sec:AIC}}

The most commonly used model selection\footnote{See \cite{burnham_2002_model} for an overview of model selection and the various techniques mentioned here.} technique is the Akaike information criterion (AIC), which arises as follows.  First, we suppose there is some true model, $M=T$, giving a distribution $\Pr(D'|T;C)$.  We quantity the discrepancy between this model and our candidate model via the \emph{Kullback-Leibler divergence}:
\begin{equation}\label{eq:KLD}
{\rm KL}(T\|M) = \mathbb E_{D'|T;C}\left[\log\frac{\Pr(D'|T,C)}{\Pr(D'|\vec{x};M,C)}\right] = \mathbb E_{D'|T;C}\left[\log{\Pr(D'|T,C)}\right] - \mathbb E_{D'|T;C}\left[\log{\Pr(D'|\vec{x};M,C)}\right],
\end{equation}
for some appropriate choice of parameters $\vec{x}$.  The ``best'' model, then, is the one that minimizes ${\rm KL}(T\|M)$, which is equivalent to maximizing the second term in \eqref{eq:KLD} since the first term only depends on the true model.  However, we do not know the true model and we do not know the best set of parameters within our candidate model.  The latter problem is naturally address by collecting data $D$ and producing an estimate of the parameters $\hat{\vec{x}}(D)$, then averaging over possible data sets such that the quantity of interest becomes
\begin{equation}
\mathbb E_{D|T;C}\left[ \mathbb E_{D'|T;C}\left[\log{\Pr(D'|\hat{\vec{x}}(D);M,C)}\right]\right].
\end{equation}
Akaike showed that, independent of the true model (and under some regularity conditions), an unbiased estimator of this quantity is
\begin{equation}\label{eq:AIC}
{\rm AIC}(M) = \max_{\vec{x}} \log\Pr(D|\vec{x};C,M) - d,
\end{equation}
where $d$ is the number of dimension of the model (the number of free parameters).  The preferred model is the one with largest value of AIC$(M)$.  The simple linear penalization with dimension makes it is clear how models with more parameters are penalized.

\subsection{Bayesian Information Criterion (BIC)\label{sec:BIC}}
The Bayesian approach is more general.  Being so, it is less obvious how it might penalize complex models.  Here we show how an asymptotic approximation leads to a form similar to the AIC.
First, we write the marginal likelihood \eqref{marginal likelihood} as the integral expectation
\begin{equation}\label{eq:marginal integral}
\Pr(D|M;C) = \int d\vec{x} \Pr(\vec{x}|M;C)\Pr(D|\vec{x};C,M).
\end{equation}
We will approximate this integral using Laplace's method.  To this end, consider the Taylor expansion of the log of the likelihood function about its peak (note that for brevity, we have dropped the context $C$ and model $M$ from the conditionals):
\begin{equation}
l(\vec{x}) := \log \Pr(D|\vec{x})\approx l(\vec{x}_0) +\nabla^{\rm T}  l(\vec{x}_0) (\vec{x}-\vec{x}_0) +\frac12(\vec{x}-\vec{x}_0)^{\rm T} \nabla \nabla^{\rm T} l(\vec{x_0})(\vec{x}-\vec{x_0}).
\end{equation}
If we assume the number of measurements $N\to\infty$, the law of large numbers gives us
\begin{align}
 \nabla \nabla^{\rm T} l(\vec{x_0}) &= \sum_{j=1}^N  \nabla \nabla^{\rm T} l_j(\vec{x_0}),\\
&\to \sum_{j=1}^N  \mathbb E[\nabla \nabla^{\rm T} l_j(\vec{x_0})],\\
&=-\sum_{j=1}^N I_j(\vec{x}_0),\\
&= - N I(\vec{x}_0),
\end{align}
where $I_j$ is the Fisher information of the $j$'th measurement and $I$ is the arithmetic average of these values.
Then, the integral \eqref{eq:marginal integral} becomes
\begin{align}
\Pr(D|M;C) &= \int d\vec{x} \exp(l(\vec{x}))\Pr(\vec{x}),\\
& =  \exp(l(\vec{x}_0))\Pr(\vec{x}_0)\left(\frac{2\pi}{N}\right)^{\frac d 2}|I(\vec{x}_0)|^{-\frac12}\left(1+O\left(\frac1 N\right)\right).
\end{align}
Now we take the logarithm to obtain
\begin{equation}
\log \Pr(D|M;C) \approx l(\vec{x}_0) - \frac d 2\log N +\frac d 2\log 2\pi+ \log\Pr(\vec{x_0}) +\log |I(\vec{x}_0)|^{-\frac12}.
\end{equation}
If we ignore the terms not changing with $N$, we have a new quantity
\begin{equation}
BIC(M)= \max_{\vec{x}} \log\Pr(D|\vec{x};C,M) - \frac d 2\log N,
\end{equation}
which is the well-known Bayesian information criterion or BIC.  Notice the striking similarity to the AIC \eqref{eq:AIC}.  Being nearly equivalent, the BIC is often considered in addition to the AIC.  Next, we will consider the full solution, which will allow us to obtain more accurate estimates of parameters averaged over the competing models.

Recent proposals have used the AIC/BIC on both simulated \cite{lougovski_2009_characterizing, lin_2011_information, schwarz_2013_error, vanenk_2013_when} and experimental data \cite{usami_2003_accuracy,guta_2012_rank}.  In \cite{ vanenk_2013_when}, however, the authors caution its unattended use.  The argument against AIC for quantum states, for example, is simple.  The AIC is derived from a metric which measures their closeness of model in their \emph{predictive} probability---a certainly well-motivated measure.  However, such a measure is only useful if all future measurements will be the same as those used to perform the data analysis.  That is, one can measure copies of a quantum system in some fixed set of bases, estimate the state, then use that estimate to predict the outcome of a measurement in a new basis.  Thus, in quantum theory, one ought to consider a measure on predictive distribution as maximized over all possible future measurements.  As suggested in \cite{ vanenk_2013_when}, such a measure might well be the quantum relative entropy, for example.  Here we avoid these problems by considering the full Bayesian solution, which we describe next.

\section{Bayesian model selection and averaging}\label{sec:Bayes}

\subsection{Model average estimate}

  Within the Bayesian framework, the model selection approach is no different than for parameter estimation.  Rather than focus on the distribution $\Pr(\vec{x}|D;C,M)$, we first consider $\Pr(M|D;C)$.  Using Bayes rule we have
\begin{equation}\label{eq:metaBayes}
\Pr(M|D;C)= \frac{\Pr(D|M;C)\Pr(M|C)}{\Pr(D|C)}.
\end{equation}
Often, practitioners of Bayesian methods go one step further and compare two models---say, $M_1$ and $M_2$---by taking the ratio of these posteriors,
\begin{equation}
\frac{\Pr(M_1|D;C)}{\Pr(M_2|D;C)}= \frac{\Pr(D|M_1;C)\Pr(M_1|C)}{\Pr(D|M_2;C)\Pr(M_2|C)},
\end{equation}
noticing the normalization factor cancels.  This quantity is called the \emph{posterior odds ratio} and first considered by Jeffreys \cite{jeffreys_1998_theory}.  Clearly, if the posterior odds ratio is larger than 1, we favor $M_1$.  The last fraction is called the \emph{prior odds ratio} and the unbiased choice favoring neither model is set this term equal to 1.  This leaves us with 
\begin{equation}
\frac{\Pr(M_1|D;C)}{\Pr(M_2|D;C)}= \frac{\Pr(D|M_1;C)}{\Pr(D|M_2;C)},
\end{equation}
which is called the \emph{Bayes factor} \cite{kass_1995_bayes}.  Each quantity in the ratio is marginal likelihood \eqref{marginal likelihood} of its respective model.  

For a discrete number of hypothetical models $\{M_k\}$, and assuming one model must be chosen, the optimal strategy is to compute the marginal likelihood if each model and select the one with the highest value.  Model selection of this type has been used in quantum mechanical problems of Hamiltonian finding \cite{wiebe_2014_quantum} and estimating error channels from syndrome measurements \cite{qecc paper}.

Here we will investigate the idea of using a meta-model, which is an average over those in $\{M_k\}$.  Assume that we are interested in some subset of parameters $\vec{y}$ common to all models.  Given we have taken data $D$ and computed each marginal likelihood $\Pr(M_k|D;C)$, we define the model average estimate (MAE) as
\begin{equation}\label{eq:MAE}
\hat{\vec{y}}_{\rm MAE}(D;C) = \mathbb E_{M_k|D;C}[\mathbb E_{\vec{y}|D;C,M_k}[\vec{y}]].
\end{equation}
In words, this is the average (over models) of the average (over parameters within models).  Variants of this approach are referred to as Bayesian model averaging \cite{hoeting_1999_bayesian}.

Before moving to our examples, a few comments are in order.  First, it is not necessary that the models are ``nested'' in the sense that we can order them into supersets.  The only requirement is that the parameters of interest are included in each model.  The other parameters in each model we might call \emph{nuisance parameters}.  Part of the appeal of the Bayesian approach is that these parameters are automatically dealt with and we can focus on those parameters which are of immediate interest.  Of course, the nuisance parameters can be inferred as well.

Let us wax philosophical for a moment.  What is being proposed is to select a meta-model, an average over many different physical models.  This may seem awkward for physical theories---after all, there is only one true model, right?  Not in our view.  Models are human constructs, those Platonic ideals which describe another world, a world we were clever enough to find through our mastery of mathematics and abstraction.  Here, we have dropped the idea that the point is to find the capital-T-Truth.  Rather, we measure our understanding of Nature through our ability to predict and control its behavior.  By averaging physical models, we can show that this idea has merit.

\subsection{Sequential Monte Carlo}\label{sec:SMC}

In practice, the Bayesian update rule and the expectations required in the equations above are analytically and computationally intractable since they involve complicated integrals over multidimensional parameter spaces which may include solutions to equations of motions which are themselves intractable.  To perform the calculations we turn to Monte Carlo techniques.  Our numerical algorithm fits within the subclass of Monte Carlo methods called \emph{sequential Monte Carlo} (SMC) or \emph{particle filtering} \cite{doucet_2000_on}. 

The SMC procedure prescribes that we approximate the probability distribution by a weighted sum of Dirac delta-functions,
\begin{equation}
\Pr(\vec{x}) \approx \sum_{j=1}^{n} w_j  \delta(\vec{x} - \vec{x}_j),
\end{equation}
where the weights at each step are iteratively calculated from the previous step via
\begin{equation}\label{eq:smc update}
w_j \mapsto \Pr(D|\vec{x}_j) w_j,
\end{equation}
followed by a normalization step.  The elements of the set $\{\vec{x}_j\}_{j=0}^{n}$ are called \emph{particles}.  Here, $n = {|\{\vec{x}_i\}|}$ is the number of particles and controls the accuracy of the approximation.  Like all Monte Carlo algorithms, the SMC algorithm approximates expectation values, such that
\begin{equation}
\mathbb{E}_{\vec{x}}[f(\vec{x})] \approx \sum_{j=1}^n w_j f(\vec{x}_j).
\end{equation}
In other words, sequential Monte Carlo allows us to efficiently compute multidimensional integrals with respect to the measure defined by the probability distribution.

The resultant posterior probability provides a full specification of our knowledge.  However, in most applications, it is sufficient---and certainly more efficient---to summarize this distribution.   In our context, the optimal single parameter vector to report is the mean of the posterior distribution
\begin{equation}
\hat{\vec{x}} = \mathbb{E}_{\vec{x}}[\vec{x}] = \sum_{j=1}^{n}w_j \vec{x}_j.
\end{equation}
The SMC approximation can also provide efficient calculation and description of \emph{regions} \cite{granade_robust_2012,ferrie_high_2013}.  For our purpose, we also require the SMC approximation to give an accurate and efficient esimate of the marginal likelihood Eq.~\eqref{eq:marginal integral}, which we need to calculate Bayes' rule at the level of models Eq.~\eqref{eq:metaBayes}.  Via the SMC approximation, the integral expectation in the definition of the marginal likelihood, Eq.~\eqref{eq:marginal integral}, is
\begin{equation}
\Pr(D) \approx \sum_{j=1}^n w_j \Pr(D|\vec{x}_j).
\end{equation}
It is not immediate obvious but is easy to see in hindsight that this is exactly the normalization that must be computed already in the SMC algorithm after the weight update, Eq.~\eqref{eq:smc update}, is applied.  By storing this value, we can apply Bayes rule at the meta-level of models Eq.~\eqref{eq:metaBayes}.

An iterative numerical algorithm such as SMC requires care to ensure stability.  Conditions for stability of the algorithm and the specifications of an implementation have been detailed elsewhere \cite{granade_robust_2012, wiebe_hamiltonian_2013}.  The SMC algorithm has now been used in many quantum mechanical parameter estimation problems \cite{chase_2009_singleshot,huszar_adaptive_2011, granade_robust_2012, ferrie_high_2013, wiebe_hamiltonian_2013, wiebe_2014_quantum} and a software implementation (the one used here) is available as a Python package \cite{qinfer,scipy}.

\section{Examples}\label{sec:examples}

\subsection{Rank selection\label{sub:rank}}

We consider first an example similar to Guta, Kypraios and Dryden \cite{guta_2012_rank}: $n$ qubits subjected to random Pauli measurements where the models under consideration are those of differing rank.  Each model will be denoted $M_r$ where $r$ is the rank of the unknown quantum state so that the dimension of model $M_r$ is $d = 2^{n+1}r$. We generate unknown quantum states with fixed rank $r$ as follows \cite{zyczkowski_2001_induced}.  Begin with a matrix $X\in \mathbb C^{2^n \times r}$ where each component $x_{ij}$ is chosen independently according to $\Re(x_{ij})\sim \mathcal N(0,1)$ and $\Im(x_{ij})\sim \mathcal N(0,1)$---standard Normal distributions.  Then define the rank $r$ density operator 
\begin{equation}\label{eq:X2rho}
\rho = \frac{X X^\dag}{\Tr(X X^\dag)}.
\end{equation}
If $r= 2^n$, this construction is equivalent to a Hilbert-Schmidt random density matrix.  The model parameters will be the vectorization of the matrix $X$: $\vec{x}= {\rm vec}(X)$.

We label the single qubit Pauli operators $\{\sigma_0,\sigma_1,\sigma_2,\sigma_3\}$ and the multi-qubit Paulis by
\begin{equation}
\sigma_k = \sigma_{k_1}\otimes\sigma_{k_2}\otimes\cdots\otimes\sigma_{k_n},
\end{equation}
where $k = k_1+ 4 k_2 + 4^2 k_3 + \cdots + 4^{n-1} k_n$.  Since each Pauli is idempotent, $\sigma_k^2= \mathbbm 1$, each individual measurement has $2$ possible outcomes which we label $d \in \{+1,-1\}$ for the $+1$ and $-1$ eigenvalues.  Then the likelihood function of a single measurement can be related to the expectation value via: $\Pr(\pm 1|\sigma_k)=(1\pm \langle \sigma_k\rangle)/2$.  Using the properties of the vec operation, we can write this as an explicit function of $\vec{x}$:
\begin{equation}
\Pr(\pm 1|X,\sigma_k) = \frac12\left(1\pm \frac{\Tr(X X^\dag \sigma_k)}{\Tr(X X^\dag)}\right)= \frac12\left(1\pm \frac{\vec{x}\cdot (\sigma_k\otimes \mathbbm{1})\vec{x}}{\vec{x}\cdot \vec{x}}\right) =\Pr(\pm 1|\vec{x},\sigma_k).
\end{equation}

We label the parameter vector within the rank $r$ model $M_r$ by $\vec{x}_r$ and the associated density matrix given by Eq.~\eqref{eq:X2rho} $\rho_r$.  Within each model, there is not a one-to-one correspondence between $\vec{x}_r$ and $\rho_r$---different vectors will yield the same density matrix.  Since we will be interested in obtaining accurate estimates of density matrices as quantified by some norm on the space of density matrices, we will average the models over their $\rho_r$'s rather than their $\vec{x}_r$'s.

Within each model, then, we have the mean density matrix (recall $D$ is data, $C$ is the context---that is, which Paulis were measured)
\begin{equation}
\hat\rho_r(D;C) = \mathbb E_{\vec{x}_r|D;C}\left[\frac{X_r X_r^\dag}{\Tr(X_r X_r^\dag)}\right].
\end{equation}
Explicitly, the MAE in Eq.~\eqref{eq:MAE} is 
\begin{equation}
\rho_{\rm MAE}(D;C) = \sum_r \hat\rho_r(D;C) \Pr(r|D;C). 
\end{equation}
We assume there is a true density matrix $\rho_{\rm t}$ and we judge each estimate of the true state $\hat \rho$ by its spectral distance to $\rho_{\rm t}$:
\begin{equation}
\|\rho_{\rm t} - \hat\rho\| = \sigma_{\rm max}(\rho_{\rm t}-\hat\rho),
\end{equation}
where $\sigma_{\rm max}$ denotes the largest singular value.  This is the norm induced by the usual Euclidean norm on vectors.

\begin{figure}\centering
  \includegraphics[width=\columnwidth]{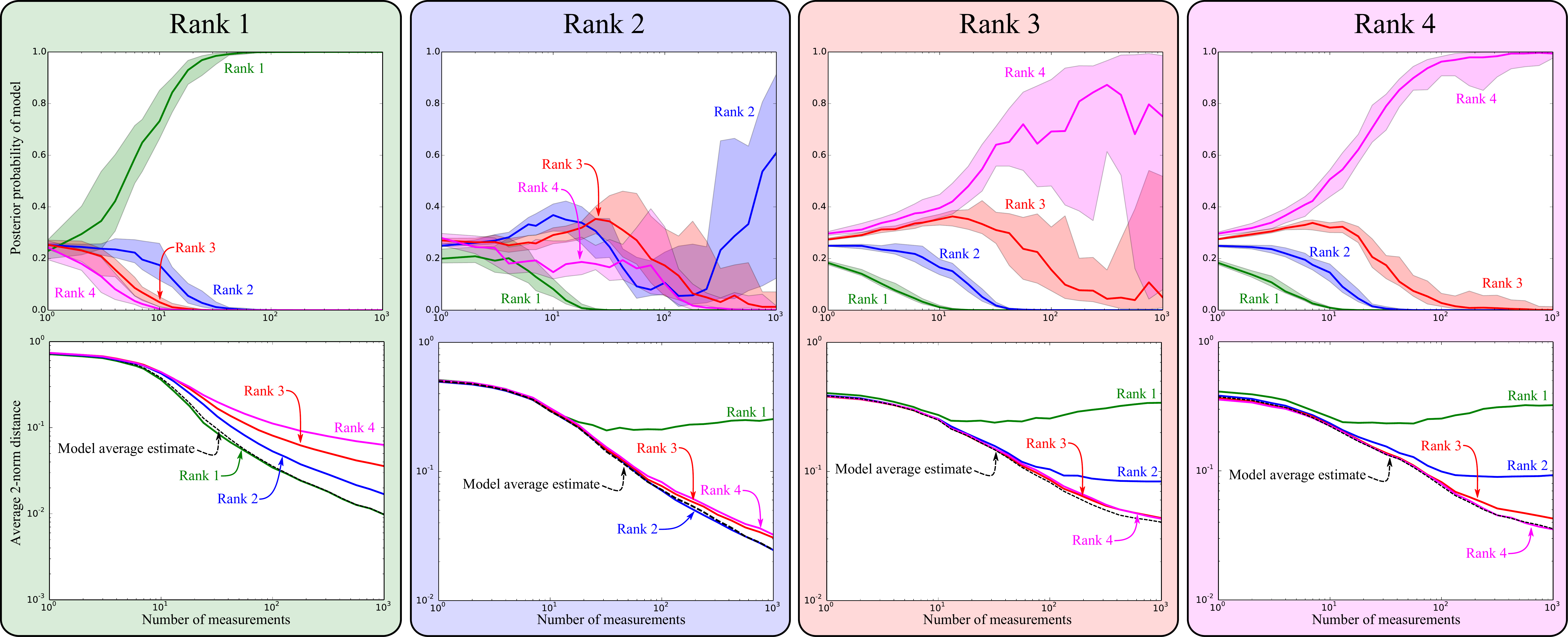}
  \caption{\label{fig:rank}  The performance of Bayesian model selection and the model average estimate for 2 qubits.  Each box represents the data for its labeled rank as the ``true'' model---from left to right, ranks 1 through 4.  The lines represent the median of the data and, where present, the shaded areas are the interquartile ranges.  Each ``measurement'' on the horizontal axis corresponds to 100 experiments of a randomly chosen Pauli measurement.  In the SMC algorithm, $10^4$ particles were used in each model.  For each true rank, 100 simulated states were generated and measured.}
\end{figure}

The data for 2 qubits is presented in Fig.~\ref{fig:rank}.  The important take-away is that the model average estimate does as well, or slightly better than the true model in every case.  Also, we see that it is quite easy to identify rank 1 models (pure states) as well as rank 4 models (full rank states) while it seems difficult to identify states of non-extreme rank.  Notice that it is extremely difficult to distinguish the rank 3 model from the full rank model when the former defines the underlying truth.  However, both models perform well with respect to the error in the estimated parameters and the model average estimate does best, on average.

\begin{figure}\centering
  \includegraphics[width=.75\columnwidth]{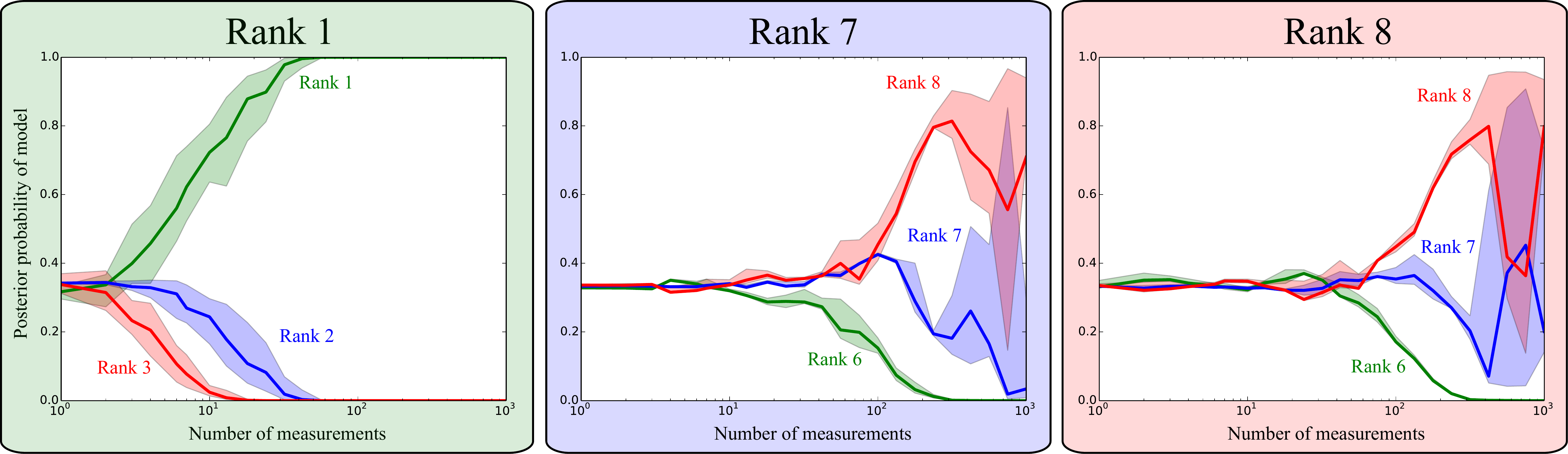}
  \caption{\label{fig:rank3}  The performance of Bayesian model selection and the model average estimate for 3 qubits.  Each box represents the data for its labeled rank as the ``true'' model.  The lines represent the median of the data and, where present, the shaded areas are the interquartile ranges.  Each ``measurement'' on the horizontal axis corresponds to 100 experiments of a randomly chosen Pauli measurement.  In the SMC algorithm, $10^5$ particles were used in each model.  For each true rank, 10 simulated states were generated and measured.}
\end{figure}

To further illustrate the difficulty in differentiating high rank states, the probabilities assigned to 3 qubit models is shown in Fig.~\ref{fig:rank3}.  Again, we see that pure states and full rank states are correctly identified, yet it is difficult to correctly distinguish between rank 7 and rank 8 states, in the same way as it was difficult to distinguish rank 3 and rank 4 states for 2 qubits.  We conclude that for the models considered here, it is easiest to correctly identify low rank and full rank states, while it is difficult to correctly identify nearly high rank states.

Notice also that models which are far away---in the sense that the ranks differ by relatively large amounts---are quickly ruled out.  In these cases, since the SMC algorithm can be run online (in parallel with the experiment), simulating such models can be stopped to mitigate the computational difficultly in simultaneously simulating many quantum models.  Still, tomography is at one extreme in the spectrum of methods for estimating quantum mechanical parameters---it is the \emph{most complete} description of the physical system.  At the other extreme is summarizing information from experiments into a single number, such as fidelity.

\subsection{Randomized benchmarking\label{sub:rb}}

In the second example, we consider the experimental protocol of randomized benchmarking \cite{magesan_2011_scalable} which has been demonstrated in a variety
of experimental settings
\cite{ryan_randomized_2009,chow_randomized_2009,olmschenk_randomized_2010,brown_single-qubit-gate_2011} to efficiently characterize noise and quantum channels.  The protocol consists of stringing together potentially long sequences of gates which is then undone to determine if the initial state has survived.  In  \cite{magesan_2011_scalable} the approach was shown to give, in expectation, an exponentially decay $A p^m + B$, where $A$ and $B$ encode the errors in preparation and measurement, $p$ is the bare survival probability and $m$ is the length of the sequence.  In the models we consider, $p$ can be related to the average fidelity over a group of gates, but more specialized protocols exists \cite{moussa_2012_practical, magesan_efficient_2012}.

Typically, $p$ is the only parameter of interest since it is directly related to the average fidelity of the device which is then compared to some threshold.  The other parameters are often consider nuisance parameters.  In \cite{granade_2014_accelerated} it was shown that the decay can be interpreted as probabilistic model where the binary outcome of each measurement sequence of length $m$ has probability of survival (labeled $0$)
\begin{equation}
\Pr(0|\vec{x}_0;m,M_0) = A_0 p^m + B_0,
\end{equation}
where the subscripts refer to this as the \emph{zeroth order model}.  In \cite{magesan_2011_scalable}, a hierarchy of models was introduced because the zeroth order model assumes the errors in the gates within the sequence are independent.  By dropping this assumption, a richer set of noise models can be studied \cite{magesan_characterizing_2012}.  Here we will study the zeroth order model and the \emph{first order model},
\begin{equation}
\Pr(0|\vec{x}_1;m,M_1) = A_1 p^m + B_1 + C_1 ( m-1)(q_1 - p^2) p^{m-2},
\end{equation}
where $A_1$ and $B_1$ again encode the preparation and measurement errors, $C_1$ encodes the error on the final gate in the sequence and $q_1-p^2$ is a measure of the gate dependence in the errors.

For the zeroth order model, we take the prior to be a normal 
 distribution with a mean vector $(p, A_0, B_0) = (0.95,0.3,0.5)$ and equal diagonal covariances given by a deviation of $\sigma= 0.01$.  Note that the first order model is equal to the zeroth order model when either $C_1=0$ or $q_1 = p^2$.  In order to not make the model so different that it would be trivial to distinguish them, we look at two priors for the first order model which are close to the zeroth order model.  The first is a normal 
 distribution with a mean vector $(p, A_1, B_1,C_1,q_1) = (0.95,0.3,0.5,0.03,0.95)=: \mu_{\rm I}$  and equal diagonal covariances given by a deviation of $\sigma= 0.01$ and the second slightly closer with the same covariance matrix but mean vector $(p, A_1, B_1,C_1,q_1) = (0.95,0.3,0.5,0.02,0.92)=:\mu_{\rm II}$.  Note that the difference between these two distribution in the relative entropy divergence is only 0.050, and so we might expect them to behave the same.  Since the both models are close to the zeroth order model, we expect it to be difficult to distinguish them.

\begin{figure}\centering
  \includegraphics[width=1\columnwidth]{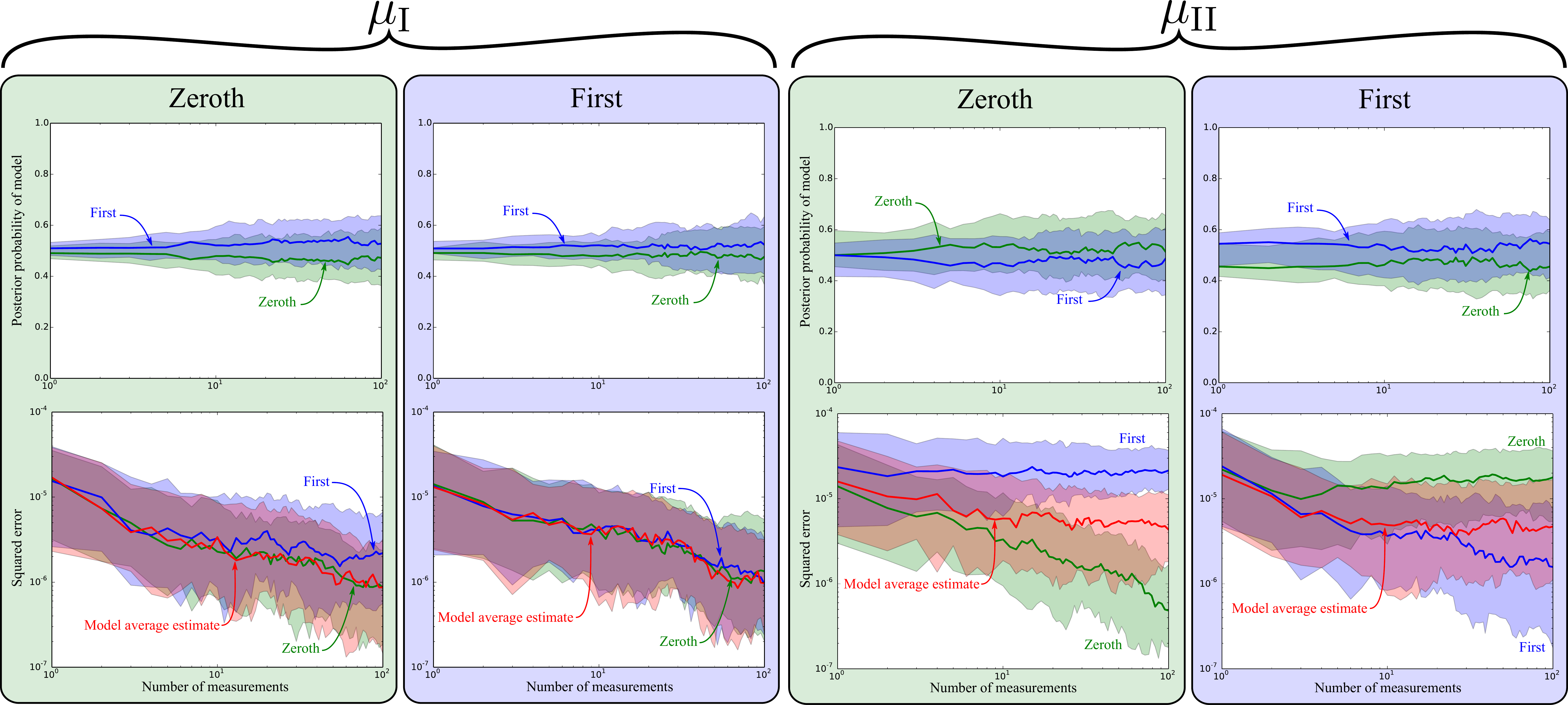}
  \caption{\label{fig:rbmodels}  The performance of Bayesian model selection and the model average estimate for the survival probability in randomized benchmarking experiments.  Each box represents the data for its label as the ``true'' model.  The lines represent the median of the data and, where present, the shaded areas are the interquartile ranges.  Each ``measurement'' on the horizontal axis corresponds to a randomized benchmarking experiment with sequences lengths $\{10,30,50,\ldots, 200\}$ and $1000$ repetitions per sequence length.  In the SMC algorithm, $10^3$ particles were used in for each model. }
\end{figure}

In Fig.~\ref{fig:rbmodels} we simulate the models noting again that, via the priors, they are very close.  This intuition is quantified by the fact that the models are hard to distinguish, regardless of which is true.  On the left in Fig.~\ref{fig:rbmodels}, we see that parameters in the first order model are so close to those in the zeroth order model that it is irrelevant which is chosen, for the purpose of estimating $p$.  However, what is ``close'' can be deceiving as we see in the right of Fig.~\ref{fig:rbmodels}.  Recall that relative entropy from $\mu_{\rm I}$ to $\mu_{\rm II}$ is only 0.050 (which explains why they are so difficult to distinguish).  In this case, the accuracy of the estimates of the parameter $p$ depend crucially on which model is actually correct.  
In such cases, the model average estimate can be seen as providing a more conservative estimate of the average gate fidelity by hedging what is at best a 50/50 guess on which model is correct.

\section{Conclusion and discussion}\label{sec:conclusion}

We have introduced a Bayesian model averaging approach to estimating parameters in quantum mechanical models describing data.  In the examples considered, the model average estimate performance as well as the unknown true model in most cases.  In situations where models are difficult to distinguish, the model average estimate can slightly \emph{outperform} the true model.

For the quantum state estimation example (Sec.~\ref{sub:rank}) we considered models of differing rank of the density matrix.  Ranks which differ by large amounts from the true rank are rapidly ruled out---that is, the probability assigned to them quickly approaches zero.  Thus, they contribute nothing to the model average estimate.  On the other hand, ranks which are close to the true rank---especially when the true rank is high---are not so easily distinguished.  This means that first selecting a rank and then performing estimation within that model can lead to overconfident estimates of the state.  By averaging, we can allow those estimates to only contribute with the relative probability that we deem them to be true.  

The mechanism for why higher rank states are hard to distinguish is yet unclear.  Although the SMC algorithm and the implementation used here have been extensively studied for a wide range of problems, it is possible that state tomography is an outlier.  That is, some major modification to modeling or the algorithm itself may be required to obtain the best performance.  This resolution would be less interesting than a physical explanation, such as Pauli measurement tomography (which we have used in the example here) is not the optimal scheme to distinguish rank.  These questions are left for future work.

In the example of randomized benchmarking (Sec.~\ref{sub:rb}), we explored a situation where the presence of higher order perturbations were difficult to detect.  The Bayesian model selection approach accurately predicts this by assigning roughly 50/50 probability assignments to the models.  Surprisingly, the ``closeness'' of the models as measured by our ability to distinguish them does not translate into our ability to accurately estimate parameters common to each.  In some cases we can do no more than guess which is correct, yet guessing the wrong model may have disastrous consequences in our ability to accurately infer the parameters of interest.  Again, the model average estimate mitigates the risk of improperly ``guessing'' which model is correct.  

We have noted in the introduction the numerous approaches to estimation within quantum theory.  This should urge one to ask, are all of these distinct approaches necessary---is there not some unified approach?  Yes!  The Bayesian framework outlined here is remarkably powerful in its generality.  We note that Bayesian ideas have already been put to good use in quantum information theoretic and foundational problems
\cite{Schack2001Quantum,Caves2002Unknown,Caves2002Quantum,Fuchs2009Priors,Fuchs2009QuantumBayesian,Fuchs2010QBism,Fuchs2011Bayesian,Leifer2011Bayesian,Leifer2012Formulating}
as well as for tomographic and parameter estimation problems \cite{Jones1991Principles,Derka1996From, Buzek1998Reconstruction,Teklu2009Bayesian, Gill2009Conciliation, blume2010optimal, Sergeevich2011Characterization,Brivio2010Experimental,Oi2012Quantum} and experimental design \cite{Ferrie2012Adaptive,Ferrie2012How}.  Importantly,  the Bayesian algorithm can provide solutions to these problems \emph{online}, while the experiment is running, with the same software tools \cite{qinfer}.  

\begin{acknowledgements}
The author thanks Chris Granade and Robin Blume-Kohout for helpful discussions.  This work was supported in part by National Science Foundation Grant No. PHY-1212445 and by the Canadian Government through the NSERC PDF program.
\end{acknowledgements}

\end{document}